\input harvmac

\def\bfone{\relax{\rm 1\kern-.35em 1}}
\def\inbar{\vrule height1.5ex width.4pt depth0pt}
\def\IC{\relax\,\hbox{$\inbar\kern-.3em{\mss C}$}}
\def\ID{\relax{\rm I\kern-.18em D}}
\def\IF{\relax{\rm I\kern-.18em F}}
\def\IH{\relax{\rm I\kern-.18em H}}
\def\II{\relax{\rm I\kern-.17em I}}
\def\IN{\relax{\rm I\kern-.18em N}}
\def\IQ{\relax\,\hbox{$\inbar\kern-.3em{\rm Q}$}}
\def\us#1{\underline{#1}}
\def\IR{\relax{\rm I\kern-.18em R}}
\font\cmss=cmss10 \font\cmsss=cmss10 at 7pt
\def\ZZ{\relax\ifmmode\mathchoice
{\hbox{\cmss Z\kern-.4em Z}}{\hbox{\cmss Z\kern-.4em Z}}
{\lower.9pt\hbox{\cmsss Z\kern-.4em Z}}
{\lower1.2pt\hbox{\cmsss Z\kern-.4em Z}}\else{\cmss Z\kern-.4em
Z}\fi}

\def\nup#1({Nucl.\ Phys.\ $\us {B#1}$\ (}
\def\plt#1({Phys.\ Lett.\ $\us  {B#1}$\ (}
\def\cmp#1({Comm.\ Math.\ Phys.\ $\us  {#1}$\ (}
\def\prp#1({Phys.\ Rep.\ $\us  {#1}$\ (}
\def\prl#1({Phys.\ Rev.\ Lett.\ $\us  {#1}$\ (}
\def\prv#1({Phys.\ Rev.\ $\us  {#1}$\ (}
\def\mpl#1({Mod.\ Phys.\ Let.\ $\us  {A#1}$\ (}
\def\ijmp#1({Int.\ J.\ Mod.\ Phys.\ $\us{A#1}$\ (}
\Title{ \vbox{\baselineskip12pt\hbox{hep-th/9611090}
\hbox{HUTP-96/A051}
\hbox{OSU-M-96-25}}}
{\centerline{Geometric Engineering of N=1 Quantum Field Theories}}
\centerline{ Sheldon Katz }
 \medskip
\centerline{ Department of Mathematics}
\centerline{Oklahoma State University}
\centerline{Stillwater, OK 74078, USA}
 \medskip
\centerline{and}
\medskip
\centerline{Cumrun Vafa}
\centerline{ Lyman Laboratory of Physics}
\centerline{Harvard University}
\centerline{Cambridge, MA 02138, USA}
\bigskip
\noindent
We construct local geometric model in terms of F- and M-theory
compactification on Calabi-Yau fourfolds which lead
to $N=1$ Yang-Mills theory in $d=4$ and its reduction on
a circle to $d=3$.  We compute
the superpotential in $d=3$,  as a
function of radius, which is generated by the Euclidean 5-brane instantons.
The superpotential turns out to be the same as the potential for affine
Toda theories.  In the limit of vanishing radius
the affine Toda potential reduces to the Toda potential.
\Date{November 1996}

\newsec{Geometric Engineering}
One of the most powerful consequences of our deeper understanding
of the dynamics of string theory has been the
appreciation of the fact that gauge
dynamics can be encoded geoemetrically in the  structure of
compactifications
of type II superstrings (see e.g. \ref\kv{S. Kachru and C. Vafa,
\nup450(1995)69,
hep-th/9505105.}\ref\aspi{P. Aspinwall,\plt357(1995)329,
\plt371(1996)231.}\ref\bsvi{M. Bershadsky, V. Sadov and C.
Vafa,\nup463(1996)398,
hep-th/9510225.}\ref\kmp{S.
Katz, D. Morrison and R. Plesser,hep-th/9601108.}\ref\km{A. Klemm and P. Mayr,
hep-th/9601014.}\ref\asg{P. Aspinwall and
M. Gross, hep-th/9602118.}\ref\sixau{
M. Bershadsky, K. Intriligator,S. Kachru,
 D. Morrison, V. Sadov and C. Vafa,hep-th/9605200.}).
 Gauge groups arise through ADE singularities of geometry
(and their fibrations
\ref\llc{A.\ Klemm, W.\ Lerche and P.\ Mayr, \plt357(1995)313}\ref\vwfi{C.\
Vafa and E.\ Witten, Proceedings
of the 1995 ICTP conference on ``S-duality and
Mirror Symmetry,'' Nuc. Phys. B (Proc. Suppl. ) 46 (1996) 225.}), whereas
matter arises as loci of enhanced
singularities \ref\kav{S. Katz and C. Vafa, hep-th/9606086.}.
This not only leads to a unified
description of gravitational and gauge theory dynamics, but it also
leads directly to a deeper understanding of gauge dynamics, even
in the limit of turning off gravitational effects \ref\kklmv{S. Kachru,
A. Klemm, W. Lerche, P. Mayr and C. Vafa,
hep-th/9508155.}\ref\klmvw{A. Klemm, W. Lerche, P. Mayr, C. Vafa, N. Warner,
hep-th/9504034.}\ref\kkv{S.~Katz, A.~Klemm,
and C.~Vafa, hep-th/9609239.}.
The basic idea is to {\it geometrically engineer} the gauge symmetry
and matter content one is interested in, and then study the corresponding
theory using string techniques.  In particular in \kkv\ it was shown
how to engineer $N=2$ theories in $d=4$.  In particular if one is
interested in studying $N=2$ theories with $G=SU(N)$ with $N_f$ flavors,
one looks for a geometry where over a ${\bf P}^1$ there is an
$A_{N-1}$ singularity and where over $N_f$ points on ${\bf P}^1$ the
singularity enhances to $A_N$.  Moreover this reduces the computation
of the prepotential in the $N=2$ field theories to the question of worldsheet
instantons of type IIA strings, which is computable using (local)
mirror symmetry.  In particular the
contribution of spacetime instantons of gauge theory
to the prepotential are mapped to the growth of the number of worldsheet
instantons in a particular configuration.

The aim of this note is to initiate a study of $N=1$ theories in $d=4$
along the same lines
and their reduction to $d=2,3$.  Our aim here is to study
the case with no matter; the case with matter can be done
in a similar way, and the results will be presented elsewhere
\ref\kvo{S. Katz and C. Vafa, work in progress.}.

\newsec{$N=1$ Yang-Mills in $d=4$ and F-theory on CY 4-fold}

If one compactifies F-theory on CY 4-folds we have an $N=1$
theory in four dimensions
(see e.g. \ref\wis{E. Witten, hep-th/9604030.}\ref\met{R.
Gopakumar and S. Mukhi, hep-th/9607057}\ref\svw{S. Sethi,
C. Vafa and E. Witten, hep-th/9606122.}\ref\dgw{R. Donagi,
A. Grassi and E. Witten, hep-th/9607091.}\ref\sch{I. Brunner, M. Lynker and R.
Schimmrighk, hep-th/9610195.}\ref\mayr{P. Mayr,
hep-th/9610162.}).  This involves studying elliptically
fibered manifolds over a 3-fold base $B$, which is the `visible'
part of the space for type IIB.  The gauge symmetries are encoded
in terms of the structure of the 7-brane worldvolume which is
$R^4\times S$ where $S$ is a 4-dimensional (2-complex dimensional)
subspace of $B$. If $n$ parallel
7-branes coincide we get $SU(n)$ gauge symmetry,
which is encoded in the elliptic fibration acquiring an $A_{n-1}$ singularity.
Similarly if we develop a $D$ or $E$ singularity we obtain $SO(2n)$ or
$E_n$ (or their modding out by outer automorphisms leading to $Sp(n)$,
$SO(2n-1)$, $F_4$, or $G_2$ \asg \sixau)
gauge symmetry in four dimensions.

The situation can in general be more complicated:  We could have a
sublocus of S, consisting of a complex curve where the singularity
gets enhanced.  It could also happen that on a number of points
on that enhanced symmetry loci, the symmetry may be further
enhanced.  In such cases the study of the theory we obtain in four
dimensions is more interesting and is expected to give matter fields
or more exotic objects such as tensionless strings, as has been
shown in a similar context for compactification of F-theory
on CY 3-folds \kav \ref\bj{M. Bershadsky
and A. Johansen, hep-th/9610111.}.
In this paper we consider mainly the case without such complications.
In other words we consider the case where on $S$ we acquire some
$ADE$ singularity, and that there are no extra singualrities anywhere
on $S$.  In this way we can geometrically engineer an $N=1$ theory in
$d=4$ with $ADE$ gauge symmetry.  We can also consider outer
automorphisms to get the non-simply laced groups.

Since we have no extra singularities on $S$ we have no matter arising
in a local way on $S$.  However global aspects of $S$ can lead to matter.
The basic idea is that on $R^4\times S$
being a D-brane worldvolume, one expects \ref\bsvii{M. Bershadsky,
V. Sadov and C.
Vafa,\nup463(1996)420,
hep-th/9511222.}\ a
partially twisted topological field theory \ref\bsjv{
M. Bershadsky, A. Johansen, V. Sadov and C. Vafa, \nup448(1995)144.}\
which is twisted along $S$ but untwisted on $R^4$, with an $N=1$ in
$d=4$. The choice of the twisting is most easily determined
by the number of supersymmetries one wishes to preserve.
  Similar topological field theories arise upon compactification
of type IIA Calabi-Yau threefolds \kmp\
and heterotic strings on CY threefolds \ref\kss{S. Kachru,
N. Seiberg and E. Silverstein, hep-th/9605036.}\ref\ksa{S. Kachru
and E. Silverstein, hep-th/9608194. }.

In the case at hand the theory is an eight dimensional
theory with $N=1$ supersymmetry with gauge group $G$.  The R-symmetry
is $U(1)$.  We compactify on $S$ which has $U(2)$ holonomy.  The supercharges
can be decomposed as
$${\bf 4}_s^{\pm}\otimes {\bf 4}_s^{\pm}$$
where the first spinor is on $R^4$ and the second on $S$
and the uncorrelated ${\pm}$ refer to the chirality of the spinors.

Given that $S$ is K\"ahler we can view the spinors as sections
of ${\cal L}\otimes \Lambda^* T^*$, where ${\cal L}$ is a square
root of the canonical line bundle and $\Lambda^*T^*$ denotes all the
 antiholomorphic p-forms.

The supercharge carries $U(1)$
R-charge.  The twisting is simply using the
$U(1)$
R-symmetry to get rid of ${\cal L}$ in front and make fermions
transform in $\Lambda^*T^*$.
  This leaves us generally with
one conserved supercharge
 for a general complex surface $S$ corresponding to the constant
function 1, and agrees with what we expect for $N=1$ supersymmetry
in $d=4$.  In eight dimensions the gauge field $A$ and the complex
scalar $\phi$
in the adjoint comprise the bosonic field of the $N=1$ multiplet.
Moreover the complex scalar carries R-charge whereas the gauge bosons
are neutral under R.  Upon topological twisting the vector
field $A$ remains a 1-form, whereas the scalars now transform
as a section of the canonical line bundle on $S$.
The number of chiral fields in the adjoint
representation we get in four dimensions from
$A$ is
$h^{1,0}(S)$, which corresponds to the number of choices for the Wilson
lines we can turn on.  The number of adjoint chiral fields
 we get from $\phi$ is equal to $h^{2,0}(S)$, i.e., the number
of zero modes of the canonical line bundle\foot{
There could in general be superpotentials for these adjoint zero modes
dictated by geometry.}.  In this paper we will assume
that both of these numbers are zero so that we do not have any
adjoint matter.  We shall return to the
more general case in a future publication.

For the most part in this paper we will  assume that
we have an $ADE$ type singularity over a complex surface $S$
which has $h^{0,1}=h^{0,2}=0$ and that the singularity does
not change over $S$.  In this case we expect to have
an $N=1$ Yang-Mills theory in $d=4$ of $ADE$ gauge group without any matter.
Note that the bare gauge coupling constant in 4 dimensions is given
by
\eqn\gco{{1\over g_4^2}=V_S}
where $V_S$ denotes the volume of $S$.  This relation follows
from the fact that the gauge coupling is of order 1 in 8 dimensions, and
upon reduction on $S$ picks up the volume factor of $S$.

The infrared dynamics in this theory is expected to involve
strong coupling phenomena of confinement and gaugino condensation.
Moreover it is expected that there will be $c_2(G)$ vacua, where
$c_2(G)$ denotes the dual Coxeter number of the group $G$,
corresponding to the choice of the phase of the gaugino condensation.
Even though it may at first sight appear difficult to see these
in this geometrical setting,
it turns out to be very easy once we take one of the dimensions
of space to be a compact circle of radius $R$.  We return
to these issues after we discuss the compactification on a circle to 3
dimensions.

\newsec{$N=2$ Yang-Mills in $d=3$ and M-theory on CY 4-fold}
If we compactify the $N=1$ theory from $d=4$ to $d=3$ we obtain
an $N=2$ theory in $d=3$.  By the chain of duality in \ref\vf{
C. Vafa, \nup469(1996)403.}\
the compactification of F-theory on a circle  is dual to M-theory on
the same elliptic Calabi-Yau where the radius of the circle $R$ is related
to the K\"ahler class of the elliptic fiber $k_E={1\over R}$. If we want
to retain the R-dependence in the physical quantities, we have
to note that the 4-fold is an  elliptic one with a singularity
over the surface $S$.

$N=2$ in $d=3$ has a Coulomb branch:  The Wilson line of the
four dimensional gauge field along the circle as well
as the dual to the vector gauge field in $d=3$ which is
a scalar, form a complex scalar field $\phi$.  Going
to
non-zero value of $\phi$ is realized geometrically by blowing
the singularity of ADE type, and $\phi$ is identified with the blow
up parameter.

For $N=2$ in $d=3$ Yang-Mills, one expects to obtain
a superpotential \ref\wah{I. Affleck, J. Harvey and E.
Witten\nup206(1982)413.}\
$W(\phi)$. In particular for the $SU(2)$ gauge group it was shown in \wah\
that a non-perturbative superpotential is generated:
$$W={\rm exp}(-{\phi\over g_3^2})$$
where $g_3$ is the 3-dimensional gauge coupling constant.
If this theory comes from a reduction of $N=1$ in $d=4 $ on a circle
of radius $R$
where $1/g_3^2=R/ g_4^2$, the superpotential develops an $R$ dependent
piece.  In fact it was argued in \ref\sws{N. Seiberg and E. Witten,
hep-th/9607163.}\
that the $R$ depenendent superpotential is (in a particular
normalization of $\phi$)
$$W={\rm exp}(-{\phi\over g_3^2})+{\rm exp}(-{1\over R g_3^2} +{\phi \over
g_3^2})$$
This in particular is consistent with the fact that for any finite $R$
there are $2 $ vacua (which solve $\partial_\phi W=0$), in agreement
with the Witten index for the $N=1$ theory in $d=4$.
We can now ask whether we can see such superpotential
directly from our geometric engineering of these field theories.
We will see that not only this is possible but it also sheds
light on the struture of the superpotential for $N=2$ pure Yang-Mills
in $d=3$.

\newsec{Generation of Superpotential}
It was shown in \wis\
how to compute the effect of zero size instantons in
M-theory and F-theory on Calabi-Yau fourfolds.
This was further studied in the context of some explicit examples
in \dgw\mayr \sch.  This method for
computation of the superpotential
applies to cases where the field theory admits a phase
where the non-abelian gauge symmetry is either completely
broken or broken to abelian parts.  This is in particular expected
to be the case for the local model of $N=2$ theories in $d=3$
we discussed above.

The general statement in \wis\ is that the superpotential
receives contributions from Euclidean 5-branes wrapped around
non-trivial 6-cycles of the Calabi-Yau fourfold. For a smooth
6-cycle $C$ a necessary condition for contribution to the superpotential
is that the holomorphic Euler characteristic of $C$ be equal to 1, i.e.
$$\chi ({\cal O}_C)=h^{0,0}-h^{1,0}+h^{2,0}-h^{3,0}=1$$
Moreover if $h^{1,0}=h^{2,0}=h^{3,0}=0$ then this is a sufficient
condition for the generation of the superpotential.

The case we have is a local model  of an ADE singularity of the
fiber over the base $S$.
We can describe the local model of this geometry as follows.  Suppose that
we have a local Calabi-Yau 4-fold $Y$ which has an ADE singularity along
$S$ and which admits a split simultaneous resolution.  By this, we mean
that there
is assumed to exist a resolution $\pi:X\to Y$ of the ADE singularity with
some additional properties.  We denote the exceptional divisor by $D$, which
is a possibly singular 3-fold.  The map $\pi$ restricts to
$D$ giving a map $\pi|_D:D\to S$ whose fibers are described
by the appropriate $ADE$ Dynkin diagram.  The fibers are allowed to
change over different points of $S$, i.e.,  we can  allow for
matter (or some exotic physics of enhanced
symmetry points).  The ``split''
assumption means that $D$ is a union of $r$ irreducible
components $D_i$, i.e., $D=\cup_i D_i$ where $r$ is the rank of the gauge
group,
i.e.\ the rank
of the singularity at the generic point of $S$.  The divisors $D_i$ are
smooth 3-folds.  Each $D_i$ can be viewed as a ${\bf P}^1$ bundle
over $S$.

If we have an elliptic fibration which degenerates over $S$ and can be
resolved as above, then the fiber
over $S$ decomposes as $D\cup D'$, where $D'$ is the closure of the
complement of the exceptional set $D$ inside the resolved elliptic fiber.
Our assumptions now says that $D'$ is a ${\bf P}^1$ bundle over $S$.
The ${\bf P}^1$ fiber corresponding to $D'$ forms the extra
node to make the Dynkin diagram an {\it affine} Dynkin diagram.
Note that this is consisent with the fact that the sum (with multiplicities)
over all ${\bf P}^1$'s in the fiber gives the class of the
elliptic curve $e$, which satisfies $e\cdot e=0$, the intersection being
taken within the fiber.
To put in more detail,
after blowing up, the elliptic fiber over $S$
will consist of intersecting spheres which form the {\it affine}
Dynkin diagram of ADE.  For example for an
$A_{n-1}$ singularity, the elliptic fiber decomposes into a cycle of $n$
spheres $e_i$ which intersect two others with intersection number one.
\foot{Type III and IV configurations are possible as well in the respective
cases of $A_1$ and $A_2$.}
For any ADE we have $r+1$ classes where $r$ corresponds to the rank of the
group. We can associate each $e_i$ with a node on the extended Dynkin diagram.
Then there is a relation \ref\math{M.~Artin, Amer.\ J.\ Math.\ {\bf 88} (1966)
129.}
\eqn\tok{\sum_{i=1}^{r+1} a_i e_i =e}
where $a_i$ denotes the Dynkin number associated with the Dynkin
node $e_i$ and $e$ denotes the class of the elliptic fiber.

Note that blowing up the singularity can only be done
in M-theory, because from the M-theory
viewpoint, equation \tok\ implies that
if the singularity is blown up the
K\"ahler class of the elliptic fiber is not zero.
The F-theory limit is obtained by pushing the K\"ahler class
of the elliptic fiber to zero size, which means
turning off the blow up modes. This is
of course expected because only in the M-theory case we have a
Coulomb
phase (corresponding to the expectation value for
additional scalar in the
gauge multiplet).  $N=1$ pure Yang-Mills theory in
4-dimensions has no moduli.

To compute the 5-brane corrections to the superpotential
we have to identify the complex 3-surfaces with holomorphic
Euler characteristic 1.  There are $r+1$ such
instantons and they are simply in one to one
correspondence with the complex 3-folds consisting
of the $e_i$ sphere over $S$.  We see this as follows.

Recall that for any complex manifold $N$, the cohomology group $H^i(N,
{\cal O}_N)$ can be identified with the $i^{\scriptstyle{\rm th}}\
\bar\partial$-cohomology group of $N$.  If there is a fibration $\phi:N\to P$
of complex manifolds $N,P$ whose fibers  are smooth and
have vanishing higher
$\bar\partial$-cohomology groups, then it is not too hard to see that the
$\bar\partial$-cohomology groups of $N$ and $P$ must coincide.  As a
consequence, we conclude in this case that $\chi({\cal O}_N)=\chi({\cal O}_P)$.

We want to apply this argument to the fibrations $D_i\to S$ and $D'\to S$
arising from our elliptic fibration.  Although $D_i,\ D'$, and $S$ are
smooth, the fibers may be singular, so the argument does not apply.  But
there is a simple generalization.  Returning to the general situation
$\phi:N\to P$, let's remove the condition that the fibers are smooth with
vanishing higher $\bar\partial$-cohomology, and replace it by the conditions
$$H^i(\phi^{-1}(p),{\cal O}_{\phi^{-1}(p)})=0\ \forall i>0,\ p\in P$$
and the condition that the only holomorphic functions on the fibers are
constants.
We again can conclude that $H^{i,0}(N)\simeq H^{i,0}(P)$ for all $i$ and
therefore $\chi({\cal O}_N)=\chi({\cal O}_P)$.
\foot{This can be established using the Leray spectral sequence~\ref\gh
{P.~Griffiths and J.~Harris, {\it Principles of Algebraic Geometry},
Wiley-Interscience, New York 1978.}
$$H^p(P,R^q\phi_*({\cal O}_{N}))\Rightarrow H^{p+q}(N,{\cal O}_{N})$$
which degenerates as our assumptions show that
$R^q\phi_*({\cal O}_{N})=0$ for $q>0$. Furthermore, our assumptions
also imply that $R^0\phi_*({\cal O}_{N})=
\phi_*({\cal O}_{N})={\cal O}_P$.
We then conclude that
$$H^i(N,{\cal O}_{N})\simeq H^i(P,{\cal O}_P).$$
for all $i$.  This implies that $h^{i,0}(N)=h^{i,0}(P)$, establishing the
desired result.}

Returning to the fibrations $D_i\to S$ and $D'\to S$, and recalling that all
of the fibers arise as exceptional curves of ADE resolutions, we only have
to compute the holomorphic cohomology groups of these exceptional curves.
The holomorphic cohomology groups for $i>1$ vanish since the fibers are curves.
It remains only to compute $H^0$ and $H^1$.

If $g:\tilde{M}\to M$ is the minimal resolution of any rank~$r$
ADE surface
singularity $p\in M$, then the exceptional curve $C=g^{-1}(p)$
is a union of $r$ copies of ${\bf P}^1$ intersecting according to the Dynkin
diagram.  The equations vanishing on $p$ pull back to equations defining $C$,
and the components $e_j$ of $C$ occur with multiplicity equal to the
corresponding Dynkin number $a_j$.  We can express this as
$C=\sum a_j e_j$.\foot{More precisely, $C$ has the structure
of a scheme. It is possible for schemes to have nonconstant holomorphic
functions on compact connected sets.  This is why the condition on the fibers
of $\phi$ above were required.}  Explicit equations which make these
muliplicities plain are given for example in the appendix of
\ref\davebirat{D.R.~Morrison, Math.\ Ann.\ {\bf 271} (1985) 415.}.
Each curve $C$ is known to be connected and rational (so that $C\cdot C=-2$),
which implies that
$H^1(C,{\cal O}_C)=0$ and that $C$ has no nonconstant holomorphic functions.

We can now conclude that $\chi({\cal O}_{D_j})=1$ and $h^{i,0}(D_j)=0$ for
$i>0$, and that $\chi({\cal O}_{D'})=1$ and $h^{i,0}(D')=0$ for
$i>0$.

In a local model for the elliptic fibration, i.e.\ a
4-fold neighborhood $X$ of $D\cup D'$ mapping to a 4-fold neighborhood of $S$,
the only irreducible
compact 3-folds contained in $X$ are precisely the $D_j$ and $D'$.  There are
exactly $r+1$ such 3-folds, and they all have $\chi({\cal O})=1$ and
vanishing higher holomorphic cohomology. Therefore
in this situation, we always get $r+1$ instantons.

We thus obtain
\eqn\supe{W=\sum_{i=1}^{r+1} exp(-V_i)}
where $V_i$ is the complexified K\"ahler class corresponding
to the $S_i$ which is the total space of $e_i$ over $S$.
For $i=1,...,r$ this can be viewed as
$$V_i=V_S V_{e_i}={1\over g_3^2} \phi_i$$
where we identify the blow up paramater $\phi_i$ with the
volume of $e_i$.  In the limit where the K\"ahler class
of the elliptic fiber becomes large, the relevant part of the
singularity is captured by the ordinary Dynkin diagram.  Note that
the contribution of the
extra class corresponding to the extra node
of the Affine Dynkin diagram to the superpotential
can be computed from \tok . Note that the volume
of the Euclidean fivebrane corresponding
to the extended node is $V_S V_{T^2-\sum_i a_i e_i}$and gives
rise to the superpotential
$$
 {\rm exp}(-V_SV_{T^2-\sum a_ie_i})={\rm exp}(-V_SV_{T^2})
{\rm exp}(V_{\sum a_ie_i})={\rm exp}(-{1\over Rg_3^2}){\rm exp}(\sum a_iV_i)$$
where we used the relation between the radius $R$ of F-theory
compactification and the
K\"ahler class of the $T^2$ fiber in the M-theory.  Thus putting
the contribution of all the $r+1$ instantons together we find
\eqn\finf{W=\sum_{i=1}^{r}{\rm exp}(-V_i)+\gamma {\rm exp}(\sum_{i=1}^r a_i
V_i)
}
where
$$\gamma={\rm exp}(-1/Rg_3^2)$$
For the case of $SU(2)$ the result above agrees with the results
obtained recently in \sws .  The generalization
we have found here to all the gauge
groups is extremely simple and suggestive.
In fact the superpotential \finf\
is exactly the same as the potential for affine Toda theory
for ADE; this link suggests that this theory may in some sense
be integrable.  The radius dependent factor $\gamma$ goes from
$1$ to $0$ as we go from $R=\infty$ to $R=0$.    In the $R=0$ limit
the potential for the affine Toda theory goes over to
that of the Toda theory.  Thus the connection from 3 dimensions
to 4 dimensions amounts to replacing the Toda superpotential
with the affine Toda version.  This is similar to the integrable structure
found recently \ref\nek{N. Nekrasov, hep-th/9609219}\ in the context of $N=2$
theories
in going from $d=4$ to $d=5$ in which a prepotential of an integrable
system was replaced by the relativistic version in going up one dimension.

This connection with integrable structure also suggests that
upon further compactification on a circle the theory
may go over to the {\it integrable} $N=2$ supersymmetric
theories with affine Toda superpotential, studied
in \ref\fintr{P. Fendley and K. Intriligator, \nup372(1992)533 \semi
\nup(380(1992)265.}.
It would be interesting to explore this connection.

Let us subject \finf\ to a simple test.  If we have an $N=1$
pure Yang-Mills in 4 dimensions with gauge group $G$, we expect to have
$c_2(G)$ inequivalent vacua.  Note that the limit of $d=4$ is obtained
in the above by setting $\gamma =1$.  In fact for any non-zero value
of $\gamma$ we expect to find $c_2(G)$ vacua given by the
choice of phase of the gaugino condensation.
This follows from the fact that in the presence
of an instanton there are $2c_2(G)$  gaugino zero modes
and factorization arguments (with the assumption of mass gap)
leads us to $c_2(G)$ choices for vacua depending on the
phase of the gaugino condensate $\langle \lambda^2 \rangle$.
To see if this is in accord with our results
we need to find solutions to $\partial_{V_i}W=0$ in \finf .
We obtain
$${{\rm exp}(-V_i)\over a_i}=\gamma{\rm exp}(\sum a_i V_i)=\beta $$
which implies that
$$\gamma \prod_{i=1}^{r}(a_i \beta)^{-a_i}=\beta$$
Since we have the group theoretic
identity $\sum_{i=1}^r a_i=c_2(G)-1$ this leads
to
$$\beta^{c_2(G)}= A$$
where $A=\gamma \prod_i a_i^{-a_i}$.  This equation has $c_2(G)$ solutions
for $\beta$ as expected.  In fact we can identify the
configuration of vacua we have found with the choices for discrete
't Hooft fluxes \ref\tho{G. 't Hooft, \nup153(1979)141.}
along the compact direction.  After all, $\phi_i$ represents
electric and magnetic Wilson lines along the compact direction.
 This is also in the same spirit as the computation
of the number of ground state vacua in this case \ref\witind{E.
Witten, \nup202(1982)253.}.

We would like to thank P. Cho, A. Johansen and E. Silverstein
for valuable discussions.

The research of S.K. was supported in part by NSF grant DMS-9311386 and
NSA grant MDA904-96-1-0021, and that of C.V. was supported in part by
NSF grant PHY-92-18167.

\listrefs

\end